\documentclass[superscriptaddress,onecolumn,secnumarabic,
nobibnotes,aps,prd,nofootinbib]{revtex4}
\usepackage{graphicx}
\usepackage{epsf}
\usepackage{bm}
\usepackage{amsmath}
\usepackage{amsfonts}
\usepackage{amssymb}
\usepackage{epstopdf}
\usepackage{natbib}
\usepackage{color}
\usepackage{color}
\providecommand{\U}[1]{\protect\rule{.1in}{.1in}}

\newcommand{\be}{\begin{equation}}
\newcommand{\ee}{\end{equation}}
\newcommand{\bea}{\begin{eqnarray}}
\newcommand{\eea}{\end{eqnarray}}

\begin{document}

\title{Dark Energy with Phantom Crossing and the $H_0$ tension}

\author{Eleonora Di Valentino}
\email{eleonora.di-valentino@durham.ac.uk}
\affiliation{Institute for Particle Physics Phenomenology, Department of Physics, Durham University, Durham DH1 3LE, UK.}

\author{Ankan Mukherjee}
\email{ankan.ju@gmail.com}
\affiliation{Centre for Theoretical Physics, Jamia Millia Islamia, New Delhi-110025, India.}
\affiliation{Department of Physics, Bangabasi College, Kolkata-700009, India.}

\author{Anjan A. Sen}
\email{aasen@jmi.ac.in, anjan.sen@ahduni.edu.in}
\affiliation{Centre for Theoretical Physics, Jamia Millia Islamia, New Delhi-110025, India.}
\affiliation{School of Arts and Sciences, Ahmedabad University, Ahmedabad 380009, India.}

\begin{abstract}
We investigate the possibility of phantom crossing in the dark energy sector and solution for the Hubble tension between early and late universe observations. We use robust combinations of different cosmological observations, namely the CMB, local measurement of Hubble constant ($H_0$), BAO and SnIa for this purpose. For a combination of CMB+BAO data which is related to early Universe physics, phantom crossing in the dark energy sector is confirmed at $95$\% confidence level and  we obtain the constraint $H_0=71.0^{+2.9}_{-3.8}$ km/s/Mpc at 68\% confidence level which is in perfect agreement with the local measurement by Riess et al. We show that constraints from different combination of data are consistent with each other and all of them are consistent with phantom crossing in the dark energy sector. For the combination of all data considered, we obtain the constraint $H_0=70.25\pm 0.78$ km/s/Mpc at 68\% confidence level and the phantom crossing happening at the scale factor $a_m=0.851^{+0.048}_{-0.031}$ at 68\% confidence level.
\end{abstract}

\maketitle

\section{Introduction}
The observed phenomenon of cosmic acceleration \cite{Riess:1998cb,Perlmutter:1998np} brought revolutionary change in our understanding about the cosmos. To explain the alleged accelerated expansion within the regime of General Relativity, it is essential to introduce some unknown source in the energy budget of the universe. This exotic source of energy is dubbed as {\it dark energy}. Different prescriptions from different branches of theoretical physics regarding the physical entity of dark energy are available in the literature (see~\cite{Bamba:2012cp} and references therein). The energy density of vacuum \cite{Carroll:2000fy,Peebles:2002gy,Padmanabhan:2002ji}, scalar fields energy density \cite{Copeland:2006wr}, or some unknown fluid \cite{Kamenshchik:2001cp,Bento:2002ps} can be candidate of dark energy. But none of these are beyond ambiguity. The unprecedented technical developments in cosmological observations in the recent years, like the observation of Cosmic Microwave Background (CMB) by Planck \cite{Akrami:2018vks}, the extended  Supernova Cosmology Project \cite{SuzukiSCP}, the observation of baryon distribution in the universe by Baryon Oscillation Spectroscopic Survey (BOSS) \cite{DawsonBOSS}, multi-wavelength observation of the large scale structure of the universe by Sloan Digital Sky Survey (SDSS) \cite{AlamSDSS} etc, have ensured very precise constraints on cosmological models.

\par Depending on the nature of the dark energy equation of state, the time varying dark energy models are classified into two section, phantom dark energy ($w_{de}<-1$) and non-phantom dark energy ($w_{de}>-1$). The phantom barrier is delineated by $w_{de}=-1$ which represents the cosmological constant or the vacuum dark energy. The prime motivation of the present work is to check whether cosmological observations allow a dark energy to have a transition from phantom to non-phantom or vice-versa. The theoretical background of phantom crossing dark energy are discussed in~\cite{Vikman:2004dc,Nojiri:2005sr,Nojiri:2005pu,Nojiri:2006ww,Bamba:2008hq,jaime,sarikd}, in the composite scalar field model in~\cite{Chimento:2008ws,Hu:2004kh,Wei:2005si} and in context of Horndeski's Theory~\cite{Deffayet:2010qz,Matsumoto:2017qil}. Some recent studies regarding the observational aspects of phantom crossing dark energy are referred there in~\cite{Nesseris:2006er,Fang:2008sn,Shafer:2013pxa,Wang:2018fng}. Recent model independent reconstruction of dark energy equation state by Zhao et al.~\cite{zhaoPRL,zhao} shows that a combination of cosmological data including CMB data from Planck observation, points towards possible phantom crossing in dark energy equation of state. Similar results have been obtained by Capozziello et al.~\cite{Capozziello:2018jya} with only low-redshifts data. Moreover reconstruction procedures for the dark energy density $\rho_{de}(z)$ \cite{Wang:2018fng} as well as Hubble parameter $H(z)$~\cite{Dutta:2018vmq} also exhibited phantom crossing in the dark energy sector. 

\par Another serious issue of dark energy reconstruction is the disagreement of the local measurement of Hubble parameter with the value estimated from the CMB. The local measurements suggest a higher value of the present Hubble parameter ($H_0$) compared to the value estimated for the standard model composed by a cosmological constant with cold dark matter ($\Lambda$CDM) from the CMB likelihood. The latest measurement of $H_0$, reported by the SH0ES collaboration, is $H_0=74.03\pm 1.42$ km/s/Mpc at 68\% CL \cite{Riess:2019cxk} and the value estimated by Planck for $\Lambda$CDM is $H_0=67.27\pm0.60$ km/s/Mpc at 68\% CL \cite{Aghanim:2018eyx}. The tension is now at 4.4$\sigma$ level.
There are many attempts to alleviate the issue in the literature
(see for an incomplete list of works Refs.~\cite{DiValentino:2016hlg,Bernal:2016gxb,Kumar:2016zpg,Kumar:2017dnp,DiValentino:2017iww,DiValentino:2017oaw,DiValentino:2017rcr,DiValentino:2017zyq,Sola:2017znb,Nunes:2018xbm,Yang:2018euj,Yang:2018uae,Yang:2018qmz,Poulin:2018cxd,Mortsell:2018mfj,Martinelli:2019dau,Vattis:2019efj,Kumar:2019wfs,Agrawal:2019lmo,Yang:2019qza,Yang:2019uzo,DiValentino:2019exe,Pan:2019gop,Martinelli:2019krf,Pan:2019hac,DiValentino:2019dzu,DiValentino:2019ffd,DiValentino:2019jae,Colgain:2019joh,Alcaniz:2019kah,Pan:2019jqh,Berghaus:2019cls,Knox:2019rjx,Pandey:2019plg,Adhikari:2019fvb,Hart:2019dxi,Liao:2020zko,Benevento:2020fev,Vagnozzi:2019ezj,Blinov:2020hmc,Chudaykin:2020acu,Alestas:2020mvb,Wang:2020zfv} and the recent overview in~\cite{DiValentino:2020zio,DiValentino:2021izs}). It has been recently discussed ~\cite{Alestas:2020zol,Camarena:2021jlr} that a transition in absolute magnitude $M_{B}$ for SnIa can also explain the apparent tension between the local and CMB measurements of Hubble parameter $H_{0}$. Such variation in $M_{B}$ in SnIa can be related to apparent variation of normalized Newtonian constant $\mu= G_{eff}/G_{N}$.

An important aspect of the present reconstruction is, therefore, to investigate whether a phantom crossing in dark energy evolution can alleviate the present Hubble tension. The present reconstruction is purely phenomenological based on parametrization of the dark energy density. There is no assumption about the physical entity of dark energy from any theoretical background apart from that it has a phantom crossing at some stage during its evolution. The dark energy density is parametrized using a Taylor series expansion truncated at certain order. The coefficients of the series expansion are constrained using observational data with a statistical approach. We have assumed that the components in the energy budget, namely the matter, dark energy and radiation, are independently conserved.
In the following sections, we discuss the present reconstruction, the observational constraints and finally conclude with overall remarks on the results.

\section{Reconstruction of the model}

One can parametrize the phantom crossing behaviour in the dark energy either through its equation of state $w_{DE}(z)$ or directly through its energy density $\rho_{DE}(z)$. On the hand, different observables are directly related to the Hubble parameter $H(z)$ rather than the equation of state of the dark energy fluid. If one parametrizes dark energy with $w_{DE}(z)$, the dark energy contribution in $H(z)$ involves the integration of $w_{DE}(z)$ over redshift interval, whereas parametrizing dark energy with $\rho_{DE}(z)$ contributes directly to $H(z)$. Hence $\rho_{DE}(z)$ is the simpler and more direct way to parametrize the dark energy contribution in $H(z)$. Hence we choose $\rho_{DE}(z)$ to model the dark energy behaviour.

Let us write the energy conservation equation for the dark energy fluid: $\frac{d\rho_{DE}}{d a} = - \frac{3}{a}(1+w_{DE})\rho_{DE}$. It is straightforward to see that for $(1+w_{DE}) > 0$ (non-phantom models) $\rho_{DE}$ decreases with scale factor, whereas for $(1+w_{DE}) < 0$ (phantom models) $\rho_{DE}$ increases with scale factor. For $w_{DE} = -1$, $\rho_{DE}$ is constant and that is the "Cosmological Constant". 
Hence, for any phantom crossing, dark energy density should pass through an extremum at some redshift $a= a_{m}$ where $\frac{d\rho_{DE}}{d a}$ changes its sign. We do a Taylor series expansion of $\rho_{DE}$ around this extremum at $a=a_m$,

\bea
\nonumber
\rho_{DE}(a)=\rho_0+\rho_2(a-a_m)^2+\rho_3(a-a_m)^3\\ 
 =\rho_0[1+\alpha (a-a_m)^2 +\beta (a-a_m)^3].
\label{eq:rho_exp}
\eea \label{rhoDE}
Here we normalize the present day scale factor $a_{0} =1$. As we have assumed that $\rho_{DE}$ has an extrema at $a_m$, we have ignored the first order derivative term in the Taylor expansion. We also restrict ourselves up to third order in the Taylor expansion. Allowing higher order terms will involve more parameters in the model that may not be tightly constrained with present data. One should also note that there can be a second extrema in $\rho_{DE}$ depending on the values of $\alpha$ and $\beta$. With this, the Hubble parameter can be written as,

\be
3H^2+3\frac{k}{a^2}=8\pi G[\rho_m+\rho_{\gamma}+\rho_{DE}].
\ee 

Finally we will have,
\bea
\nonumber
H^2(a)/H_0^2=\Omega_{m0}a^{-3}+\Omega_{k0}a^{-2}
+\Omega_{\gamma 0}a^{-4}
+ \\
+\left(\frac{1-\Omega_{m0}-\Omega_{k0}-\Omega_{\gamma 0}}{1+\alpha(1-a_m)^2 +\beta(1- a_m)^3}\right)
\left[1+\alpha (a-a_m)^2 +\beta (a-a_m)^3\right], 
\eea
and the dark energy equation of state

\be
w_{DE}(a)=-1-\frac{a[2\alpha(a-a_m)+3\beta(a-a_m)^2]}{3[1+\alpha(a-a_m)^2+\beta(a-a_m)^3]}.
\ee

\noindent
One can easily rewrite the above expression for $w_{DE}(a)$ to show that it represents a Pade series of order (3,3) which has a better convergence radius. Aditionally, for early times ($a\rightarrow 0$), the equation of state $w_{DE} \rightarrow -1$ shows the Cosmological Constant behaviour for the dark energy. This confirms that the dark energy equation of state is well behaved at early time without any convergence issue. It is also not difficult to verify that adding higher order terms in $\rho_{DE}$ does not change the $w_{DE} \rightarrow -1$ behaviour at early time. We should add that the different terms in the expression for $\rho_{DE}$ can be generated by non-canonical scalar fields with lagrangian ${\cal L} \propto - X^{n/(2(3+n))}$ with different values of $n$ as shown in~\cite{sen_pressure}.

Set of model parameters, $(\alpha, \beta, a_m)$ are introduced through the present reconstruction. Clearly the present model mimics the $\Lambda$CDM for $\alpha=\beta=0$. The $a_m$ is the scale factor, where the $\rho_{DE}$ has an extrema. If $a_m$ is constrained to be $a_m<1$ (we fix $a_{0}=1$ for the present day scale factor), it is a signature of transition in the nature of dark energy.
In our subsequent analysis, we assume spatially flat universe, i.e. $\Omega_{k0} =0$. We let the dark energy density $\rho_{DE}$ free to become negative, as considered by other works (see for example~\cite{Delubac:2014aqe,Poulin:2018zxs,Wang:2018fng,Dutta:2018vmq}).

\section{Methodology}

In order to constrain the DE models parameters, we make use of some of the most recent cosmological measurements available. These will be:

\begin{itemize}

\item {\bf CMB}: we consider the temperature and polarization CMB angular power spectra of the Planck legacy realease $2018$ {\it plikTTTEEE+lowl+lowE}~\cite{Aghanim:2018eyx,Aghanim:2019ame} as a baseline.\footnote{Note that there is an alternative likelihood for the Planck data, CamSpec~\cite{Efstathiou:2019mdh}, but they are consistent, as stated clearly from the Planck collaboration.}

\item {\bf R19}: we adopt a gaussian prior $H_0=74.03\pm1.42$ km/s/Mpc at 68\% CL on the Hubble constant as measured by the SH0ES collaboration in~\cite{Riess:2019cxk}.

\item {\bf BAO}: we add the Baryon Acoustic Oscillation measurements 6dFGS~\cite{Beutler:2011hx}, SDSS MGS~\cite{Ross:2014qpa}, and BOSS DR12~\cite{Alam:2016hwk}, as adopted by the Planck collaboration in~\cite{Aghanim:2018eyx}.\footnote{Note that there is an updated version of the BAO data~\cite{deMattia:2020fkb}, but we prefer to keep the combination used in the literature, for a better comparison.}

\item {\bf Pantheon}: we make use of the luminosity distance data of $1048$ type Ia Supernovae from the Pantheon catalog~\cite{Scolnic:2017caz}.

\item {\bf lensing}: we consider the 2018 CMB lensing reconstruction power spectrum data, obtained with a CMB trispectrum analysis in~\cite{Aghanim:2018oex}.

\end{itemize}

We adopt as a baseline a 9-dimensional parameter space, i.e. we vary the following cosmological parameters: the baryon energy density $\Omega_bh^2$, the cold dark matter energy density $\Omega_{c}h^2$, the ratio of the sound horizon at decoupling to the angular diameter distance to last scattering $\theta_{MC}$, the optical depth to reionization $\tau$, the amplitude and the spectral index of the primordial scalar perturbations $A_s$ and $n_s$, and, finally, the three parameters assumed in our expansion of the $\rho_{DE}$ in eq.~\ref{eq:rho_exp}, i.e. $\alpha$, $\beta$ and $a_m$. We impose flat uniform priors on these parameters, as reported in Table~\ref{tab:priors}.  

To analyse the data and extract the constraints on these cosmological parameters, we use our modified version of the publicly available Monte-Carlo Markov Chain package \texttt{CosmoMC}~\cite{Lewis:2002ah}. This is equipped with a convergence diagnostic based on the Gelman and Rubin statistic~\cite{Gelman-Rubin}, assuming $R-1<0.02$, and implements an efficient sampling of the posterior distribution using the fast/slow parameter decorrelations \cite{Lewis:2013hha}. \texttt{CosmoMC} includes the support for the 2018 Planck data release~\cite{Aghanim:2019ame} (see \url{http://cosmologist.info/cosmomc/}). Finally, since for point $\alpha=\beta=0$, the present model becomes the $\Lambda$CDM one, as it has already mentioned before, and the likelihood has singular nature as $a_m$ becomes redundant in this case, we switch back to the unmodified \texttt{CosmoMC} code for the analysis of this point, to avoid problems.

\begin{table}
\begin{center}
\begin{tabular}{c|c}
Parameter                    & Prior\\
\hline 
$\Omega_{b} h^2$             & $[0.005,0.1]$\\
$\Omega_{c} h^2$             & $[0.005,0.1]$\\
$\tau$                       & $[0.01,0.8]$\\
$n_s$                        & $[0.8,1.2]$\\
$\log[10^{10}A_{s}]$         & $[1.6,3.9]$\\
$100\theta_{MC}$             & $[0.5,10]$\\ 
$\alpha$                        & $[0,30]$\\ 
$\beta$                        & $[0,30]$\\
$a_m$                        & $[0,1]$\\
\end{tabular}
\end{center}
\caption{Flat priors for the cosmological parameters.}
\label{tab:priors}
\end{table}

%%%%%%%%%%%%%%%%%%%%%%%%%%%%%%%%%%%%%%%%%%%%
\begin{figure*}
\centering
\includegraphics[scale=0.55]{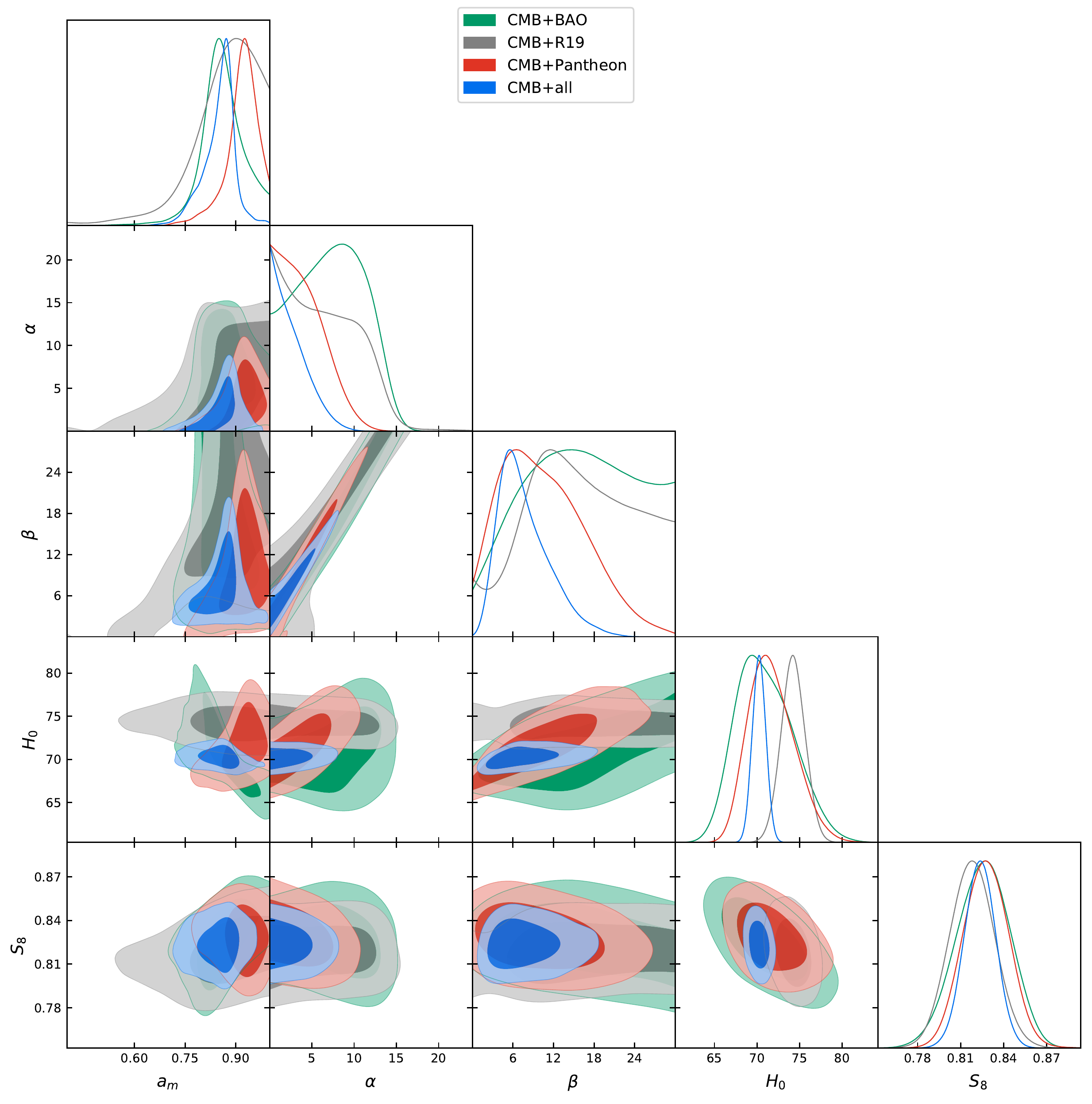}
\caption{Triangular plot showing 2D and 1D posterior distributions of some interesting parameters considered in this work. Planck+all refers to Planck+lensing+BAO+R19+Pantheon.}
\label{2D_tri}
\end{figure*}
%%%%%%%%%%%%%%%%%%%%%%%%%%%%%%%%%%%%%%%%%%%%%%%%%%%

%%%%%%%%%%%%%%%%%%%%%%%%%
\begin{table*}[tb]
\caption{{\small 68\% CL constraints on the cosmological parameters for the different dataset combinations explored in this work. CMB+all refers to Planck+lensing+BAO+R19+Pantheon.}}
\begin{center}
\resizebox{\textwidth}{!}{  
\begin{tabular}{ c |c c c c c  } 
  \hline
 \hline
 Parameters  & CMB+lensing & CMB+R19 & CMB+BAO & CMB+Pantheon & CMB+all \\ 

 \hline
  $a_m$ & $<0.276$ & $>0.830$ & $0.859\pm0.064$ & $0.917^{+0.054}_{-0.029}$ & $0.851^{+0.048}_{-0.031}$ \\
  $\alpha$ & $<17.7$ & $<8.62$ & $7.3\pm3.9$ & $<5.10$ & $<3.32$ \\
  $\beta$ & $<16.7$ & $16.0\pm7.5$ & $16.1\pm7.8$ & $10.6^{+4.4}_{-7.9}$ & $7.7^{+2.2}_{-4.7}$ \\
  $\Omega_c h^2$ & $0.1194\pm0.0014$ & $0.1196\pm0.0014$ & $0.1201\pm0.0013$ & $0.1198\pm0.0014$ & $0.1198\pm0.0011$ \\
  $\Omega_b h^2$ & $0.02243\pm0.00014$ & $0.02243\pm0.00016$ & $0.02238\pm0.00014$ & $0.02240\pm0.00015$ & $0.02240\pm0.00014$ \\
  $100\theta_{MC}$ & $1.04097\pm0.00031$ & $1.04096\pm0.00032$ & $1.04092\pm0.00030$ & $1.04095\pm0.00032$ & $1.04093\pm0.00030$ \\
  $\tau$ & $0.0521\pm0.0076$ & $0.0532\pm0.0080$ & $0.0539^{+0.0070}_{-0.0080}$ & $0.0529\pm0.0076$ & $0.0521\pm0.0075$ \\
  $n_s$ & $0.9667\pm0.0042$ & $0.9665\pm0.0045$ & $0.9652\pm0.0043$ & $0.9659\pm0.0045$ & $0.9655\pm0.0038$ \\
  ${\rm{ln}}(10^{10}A_s)$ & $3.038\pm0.015$ & $3.041\pm0.016$ & $3.044\pm0.016$ & $3.041\pm0.016$ & $3.039\pm0.015$ \\
   \hline
  $H_0 {\rm[km/s/Mpc]}$ & $>92.8$ & $74.2\pm1.4$ & $71.0^{+2.9}_{-3.8}$ & $71.7^{+2.2}_{-3.1}$ & $70.25\pm0.78$ \\
  $\sigma_8$ & $1.012^{+0.051}_{-0.009}$ & $0.881\pm0.018$ & $0.848^{+0.027}_{-0.034}$ & $0.860^{+0.026}_{-0.033}$ & $0.838\pm0.011$ \\
  $S_8$ & $0.752^{+0.009}_{-0.025}$ & $0.818\pm0.016$ & $0.826\pm0.019$ & $0.828\pm0.016$ & $0.823\pm0.011$ \\
  $r_{\rm drag}$ & $147.19^{+0.28}_{-0.26}$ & $147.14\pm0.30$ & $147.06\pm0.29$ & $147.10\pm0.30$ & $147.10\pm0.25$ \\
  
 \hline
  \hline
\end{tabular}
}
\end{center}
\label{table}
\end{table*}
%%%%%%%%%%%%%%%%%%%%%%%%%%%%%%%%%%%

%%%%%%%%%%%%%%%%%%%%%%%%%
\begin{table*}[tb]
\caption{{\small $\chi^2_{\rm bf}$s comparison between $\Lambda$CDM and Phantom Crossing for the different dataset combinations explored in this work. CMB+all refers to Planck+lensing+BAO+R19+Pantheon.}}
\begin{center}
\resizebox{0.8\textwidth}{!}{  
\begin{tabular}{ c |c c c c c  } 
  \hline
 \hline
 $\Lambda$CDM  & CMB+lensing & CMB+R19 & CMB+BAO & CMB+Pantheon & CMB+all \\ 
  \hline
  $\chi^2_{\rm bf, tot}$ & $2782.040$ & $2791.838$ & $2779.712$ & $3807.500$ & $3840.406$  \\
  $\chi^2_{\rm bf, CMB}$ & $2778.122$ & $2768.113$ & $2770.060$ & $2767.697$ & $2779.508$  \\
  $\chi^2_{\rm bf, lensing}$ & $8.981$ & $-$ & $-$ & $-$ & $9.510$  \\
  $\chi^2_{\rm bf, R19}$ & $-$ & $18.117$ & $-$ & $-$ & $16.414$  \\
  $\chi^2_{\rm bf, BAO}$ & $-$ & $-$ & $6.514$ & $-$ & $5.271$  \\
  $\chi^2_{\rm bf, Pantheon}$ & $-$ & $-$ & $-$ & $1035.268$ & $1034.768$  \\
   \hline
    \hline
 Phantom Crossing  & CMB+lensing & CMB+R19 & CMB+BAO & CMB+Pantheon & CMB+all \\ 
  \hline
  $\chi^2_{\rm bf, tot}$ & $2776.610$ & $2765.556$ & $2775.204$ & $3805.278$ & $3828.424$  \\
  $\chi^2_{\rm bf, CMB}$ & $2770.124$ & $2762.965$ & $2763.945$ & $2765.943$ & $2775.585$  \\
  $\chi^2_{\rm bf, lensing}$ & $8.145$ & $-$ & $-$ & $-$ & $8.702$  \\
  $\chi^2_{\rm bf, R19}$ & $-$ & $0.307$ & $-$ & $-$ & $8.275$  \\
  $\chi^2_{\rm bf, BAO}$ & $-$ & $-$ & $5.321$ & $-$ & $5.702$  \\
  $\chi^2_{\rm bf, Pantheon}$ & $-$ & $-$ & $-$ & $1036.603$ & $1035.971$  \\
  
 \hline
  \hline
\end{tabular}
}
\end{center}
\label{table_chi}
\end{table*}
%%%%%%%%%%%%%%%%%%%%%%%%%%%%%%%%%%%

%%%%%%%%%%%%%%%%%%%%
\begin{figure}[tb]
\begin{center}
\includegraphics[angle=0, width=0.45\textwidth]{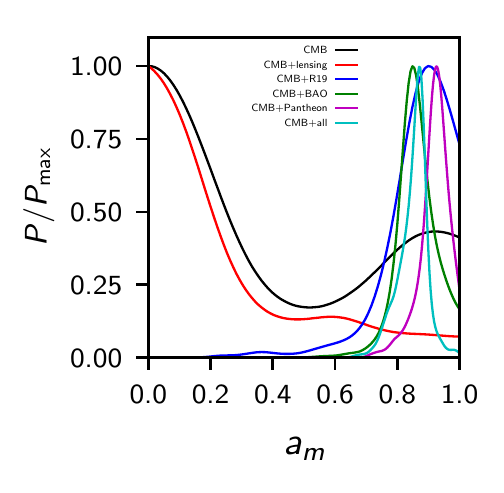}

\end{center}
\caption{{\small 1D posterior distribution of $a_m$ for the different dataset combinations explored in this work. CMB+all refers to Planck+lensing+BAO+R19+Pantheon.}}
\label{am}
\end{figure}
%%%%%%%%%%%%%%%%%%%%%%

%%%%%%%%%%%%%%%%%%%%
\begin{figure*}[tb]
\begin{center}
\includegraphics[angle=0, width=0.48\textwidth]{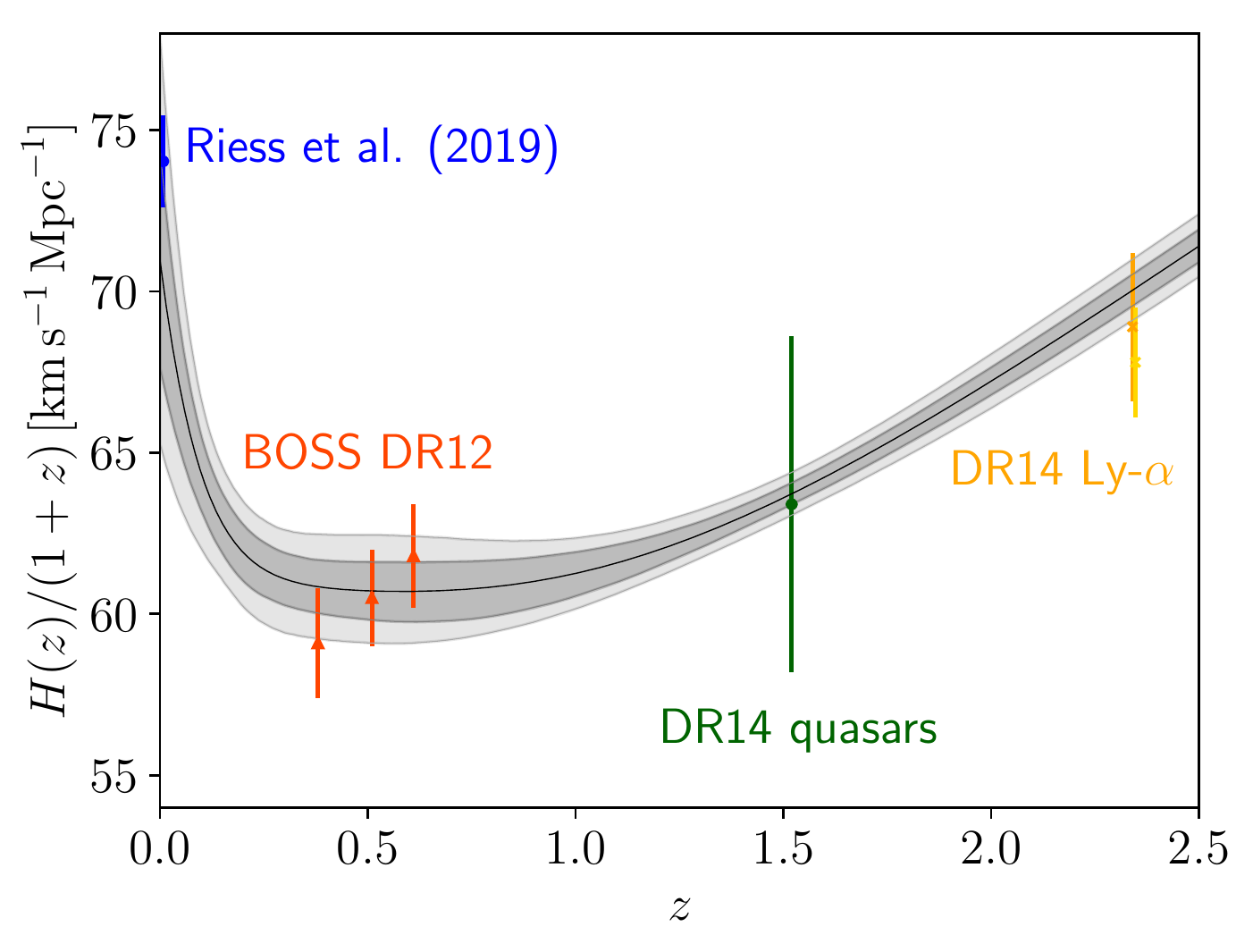}
\includegraphics[angle=0, width=0.48\textwidth]{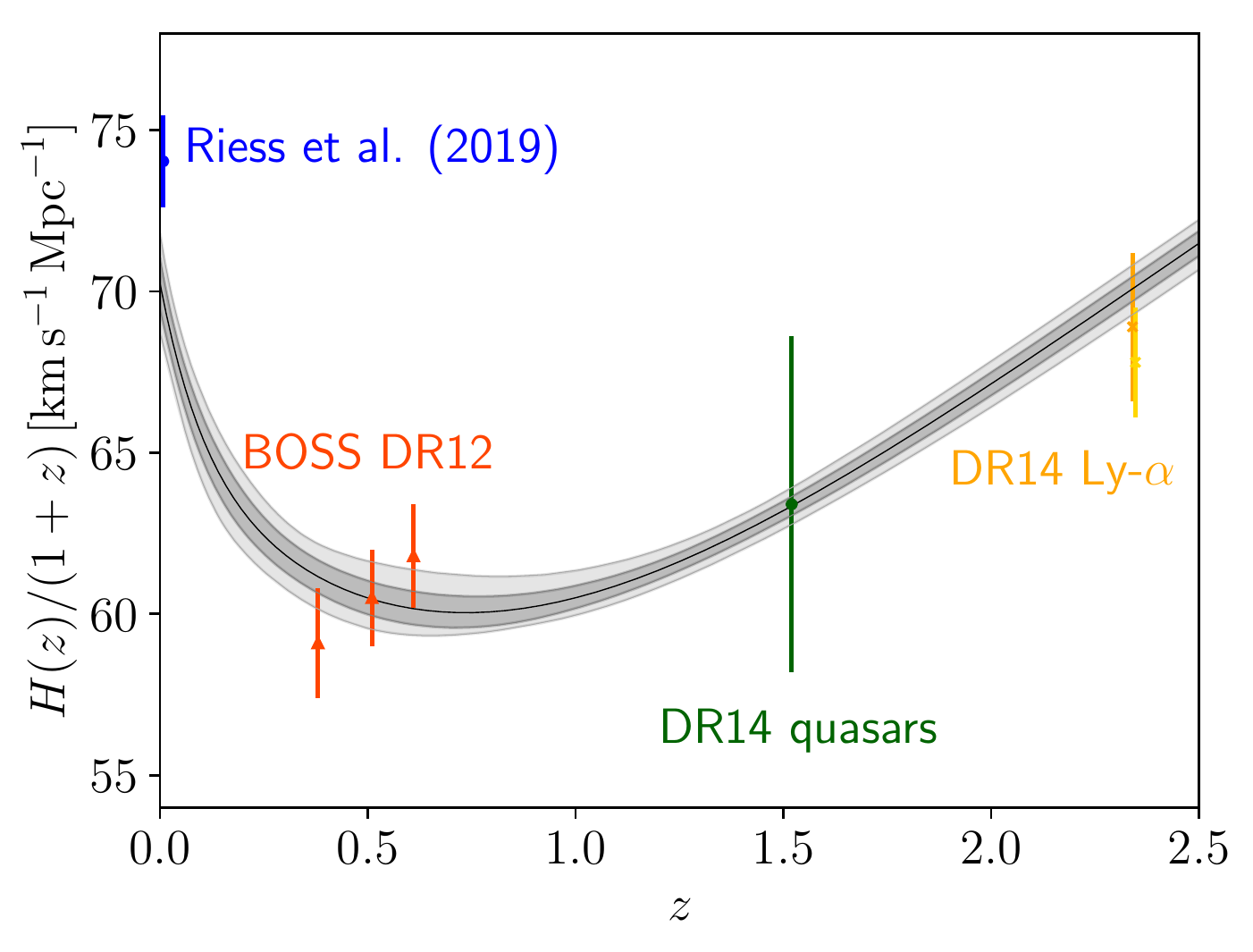}

\end{center}
\caption{\small Behaviour of $H(z)/(1+z)$ for the combinations CMB+BAO (on the left) and CMB+all (on the right). Observational data points of local measurement of $H_0$ by Riess et al.\cite{Riess:2019cxk}, BOSS DR12 \cite{Alam:2016hwk}, BOSS DR14 quasars \cite{Zarrouk:2018vwy}, BOSS DR14 Ly-$\alpha$ \cite{Agathe:2019vsu,Blomqvist:2019rah} are also shown.}
\label{fig:Hz}
\end{figure*}
%%%%%%%%%%%%%%%%%%%%%%

\section{Observational constraints}

In Table~\ref{table} we show the constraints at 68\% CL for the cosmological parameters explored in this paper, for different dataset combinations. In Fig.~\ref{2D_tri} we show instead the 2D contour plots and 1D posterior distribution on some of the most interesting parameters. We are not showing the CMB only constraints because they are bimodal in $a_m$, i.e. CMB alone is not able to distinguish which is its best value in fitting the data, but we need additional probes to broke the degeneracy. Eventually, we will find that CMB+lensing prefers one of the two peaks, while all the other combinations (+BAO, +Pantheon and +R19) prefer the other peak (see Fig.~\ref{am}). Finally, in Table~\ref{table_chi} we compare the $\chi^2_{bf}$ of the best fit of the data for the standard $\Lambda$CDM model and the Phantom Crossing. We can see that in all the combinations of data considered here, the Phantom Crossing model improves the $\Delta \chi^2$ with respect to the standard model.

Comparing the constraints on the cosmological parameters reported in Table~\ref{table} for our scenario, with those reported by the Planck collaboration in~\cite{Aghanim:2018eyx} for a $w$CDM model, we can see that they are completely in agreement for the CMB+lensing dataset combination (second column). This happens because the scale factor of the transition $a_m$ is consistent with zero, so in agreement with a phantom dark energy like preferred by Planck in a $w$CDM model. Moreover, both $\alpha$ and $\beta$ are consistent with 0, i.e. a cosmological constant, within one standard deviation.
For CMB+lensing we find that the Hubble constant parameter is almost unconstrained ($H_0>75.4$ km/s/Mpc at 95\% CL), and $S_8= 0.752^{+0.009}_{-0.025}$ at 68\% CL is completely in agreement within one standard deviation with the combination of the cosmic shear data KiDS+VIKING-450+DES-Y1~\cite{Asgari:2019fkq}, while the tension on $S_8$ is at $3.2\sigma$ in a $\Lambda$CDM context.

Since the CMB and R19 are in agreement now within 2 standard deviations, we can combine them safely together. The results we obtain for the joint analysis CMB+R19 are reported in the third column of Table~\ref{table}. Here we see that, while $\alpha$ is still consistent with zero within $1\sigma$, we have now $\beta=16.0\pm7.5$ at 68\% CL and $a_m>0.830$ at 68\% CL, i.e. consistent with $1$.

An interesting result is that obtained combining together CMB and BAO, that is shown in the forth column of Table~\ref{table}. Here we can see that, on the contrary with respect to many other cosmological scenarios, included a $\Lambda$CDM model of which our parametrization is an extension, CMB+BAO gives $H_0=71.0^{+2.9}_{-3.8}$ km/s/Mpc at 68\% CL. This large Hubble constant value is now perfectly consistent within one standard deviation with the R19 measurement, while all the other cosmological parameters are almost unchanged if compared with a $w$CDM scenario for the same CMB+BAO data combination. This increase of the $H_0$ parameter is due to its positive correlation with $\alpha$ and $\beta$, and negative with $a_m$, as we can see in Fig.~\ref{2D_tri}. For the CMB+BAO case we have in fact an indication that all these three parameters are different from the expected values at more than $1\sigma$. In particular we find, at 68\% CL, $a_m=0.859\pm0.064$, $\alpha=7.3\pm3.9$ and $\beta=16.1\pm7.8$. Therefore, in this case there is an indication at more than $2\sigma$ for a transition in the dark energy density. The constraint on the present day equation of state $w_{DE}(z=0)$ is $-1.61^{+0.60}_{-0.91}$ at $95$\% CL ruling out the cosmological constant at about 2$\sigma$. Given that both CMB and BAO are related to early Universe physics, this shows that a phantom crossing in dark energy sectors alleviates the tension between the early and late Universe determinations of the parameter $H_{0}$. In the left panel of Fig.~\ref{fig:Hz}, we show the behaviour of the expansion rate of the Universe for this dataset combination. We can see an excellent agreement with all the latest measurements. This agreement finds confirmation in the $\chi^2_{bf}$ (see Table~\ref{table_chi}), where we show that the Phantom Crossing model improves the $\Delta \chi^2$ with respect to the standard $\Lambda$CDM model, not only for the Planck+BAO combination, but also for the BAO data alone.

The same interesting larger value of the Hubble constant persists even if we combine CMB and Pantheon data. In this case, as we show in the fifth column of Table~\ref{table}, $H_0=71.7^{+2.2}_{-3.1}$ km/s/Mpc at 68\% CL, i.e. consistent with R19. As we can see in Fig.~\ref{2D_tri}, it is the positive correlation between $H_0$ and $\alpha$ and $\beta$ to shift the Hubble constant towards higher values, while, on the contrary with respect to the CMB+BAO combination, in this case there is a positive correlation also between $H_0$ and $a_m$. For the Dark Energy parameters of our model we find for CMB+Pantheon, at 68\% CL, $a_m=0.917^{+0.054}_{-0.029}$ and $\beta=10.6^{+4.4}_{-7.9}$, i.e. different from the expected value in a $\Lambda$CDM model at more than $1\sigma$, while $\alpha<5.10$ is consistent with zero. 

Given a preference for all the data combination of a large $H_0$, we can conclude that this indication is robust irrespective to the combination of data analysed here. For this reason we combine them all together because they are no more in tension. In fact, even the Planck+lensing dataset combination gives $H_0>75.4$ km/s/Mpc at 95\% CL, i.e. in perfect agreement with R19. The joint result, i.e CMB+lensing+BAO+Pantheon+R19, is displayed in the last column of Table~\ref{table}, where we see $H_0=70.25\pm0.78$ km/s/Mpc at 68\% CL, reducing the tension with R19 at $2.3$ standard deviations. In the right panel of Fig.~\ref{fig:Hz}, we show the behaviour of the expansion rate of the Universe for this combination. Also in this case, we can see a good agreement with all the latest measurements of BAO. However, even if for this dataset combination we have a slightly lower $S_8=0.823\pm0.011$ at 68\% CL, the tension with the cosmic shear data KiDS+VIKING-450+DES-Y1~\cite{Asgari:2019fkq} is still at $3.1\sigma$. For the joint case we find, at 68\% CL, $a_m=0.851^{+0.048}_{-0.031}$ and $\beta=7.7^{+2.2}_{-4.7}$, i.e. they are different from the expected value in a $\Lambda$CDM scenario at more than $2\sigma$ because highly non-gaussian, while $\alpha<3.32$ at 68\% CL is consistent with zero. Therefore, a robust indication at more than $2\sigma$ for a transition in the dark energy density is suggested by the data. The constraint on the present day equation of state $w_{DE}(z=0)$ is $-1.33^{+0.31}_{-0.42}$ at $95$\% CL, ruling out the cosmological constant at more than 2$\sigma$. Finally, if we look at the $\chi^2_{bf}$ in Table~\ref{table_chi}, we can see that the Phantom Crossing model improves significantly the total $\Delta \chi^2$ we had for the standard $\Lambda$CDM model.

From the constraints on $r_{d}$ in Table 3 for different data combinations, especially for CMB+BAO combination, we can say that we agree with BAO data, as well as larger $H_{0}$ value, even if we don't change $r_{d}$ as constrained by Planck for $\Lambda$CDM. This may be due to non-monotonic dark energy evolution in late time.

It is also not difficult to check that $\rho_{DE}(z)$ for the constrained parameter space can become negative for some redshifts and this is consistent with earlier results by \cite{Delubac:2014aqe,Poulin:2018zxs,Wang:2018fng,Dutta:2018vmq}. 
Negative $\rho_{DE}(z)$ at some earlier time may help to reduce the Hubble tension.

Additionally, to perform a model comparison, we compute the Bayesian evidence and we show the results in Table~\ref{tab:BE}. This allow us to quantify which model fits better the data between $\Lambda$CDM and Phantom Crossing.
We use the publicly available cosmological code \texttt{MCEvidence}\footnote{\href{https://github.com/yabebalFantaye/MCEvidence}{github.com/yabebalFantaye/MCEvidence}~\cite{Heavens:2017hkr,Heavens:2017afc}.}. We will have that for negative (positive) values of the Bayes factor $\ln B_{ij}$, the $\Lambda$CDM (Phantom Crossing) is the preferred model. To interpret the results, we will refer to the revised Jeffreys scale by Kass and Raftery as in Ref.~\cite{Kass:1995loi}. Therefore, we will have for $0 \leq | \ln B_{ij}|  < 1$ a Weak evidence, for $1 \leq | \ln B_{ij}|  < 3$ a Definite evidence, for $3 \leq | \ln B_{ij}|  < 5$ a Strong evidence, and for $| \ln B_{ij} | \geq 5$ a Very Strong evidence for one model versus the secon one. Looking at the Table~\ref{tab:BE} we can see that we have a Very Strong Evidence for the Phantom Crossing for CMB+R19, while the $\Lambda$CDM model is preferred for all the other dataset combinations.

%%%%%%%%%%%%%%%%%%%%%%%%%%%%%%%%%%%
\begin{table}[tb]
\caption{{\small The table shows the values of $\ln B_{ij}$ calculated for the Phantom Crossing model with respect to the $\Lambda$CDM scenario. The negative value in $\ln B_{ij}$ indicates that there is a preference for $\Lambda$CDM against the Phantom Crossing, while the positive a preference for the Phantom Crossing.}}
\begin{center}
\begin{tabular}{|c|c|c|c|c|}
\hline 
Data & $\ln B_{ij}$\\
\hline 
CMB       &       $0.30$ \\
CMB+lensing    &       $0.13$ \\
CMB+R19     &       $6.91$ \\
CMB+BAO  &     $-2.29$ \\
CMB+Pantheon      & $-4.46$ \\
CMB+all &   $-1.75$ \\

\hline 
\end{tabular}
\end{center}
\label{tab:BE}
\end{table}
%%%%%%%%%%%%%%%%%%%%%%%%%%%%%%%%%%%

\section{Conclusion}
In this work, we consider a dark energy behaviour with phantom crossing and confront it with different observational data including the latest CMB data from Planck. We do not consider any specific theoretical set up involving fields but rather we approach it in a general way where we assume that the dark energy density should have an extrema at a particular scale factor $a_{m}$ for phantom crossing. If $a_{m} < 1$, this crossing happens before the present day. We Taylor expand the dark energy density around this extrema and check whether the observational data is consistent with $a_{m} <1$. We find that a combination of observational data including that from Planck is indeed consistent with $a_{m} <1$ confirming the presence of phantom crossing. Moreover the phantom crossing also helps to alleviate the $H_{0}$ tension between low and high redshift observations. The CMB+BAO combination which represents early Universe physics, gives a constraint $H_0=71.0^{+2.9}_{-3.8}$ km/s/Mpc at 68\% CL for model with phantom crossing which is fully in agreement with the local measurement of $H_{0}$ by R19. Moreover, constraints on different parameters including $H_{0}$ for different combination of data are consistent to each other and allows us to combine all the data. For the combination of all data, the phantom crossing is observed at more than $2\sigma$ and the constraint on $H_{0}$ is $H_0=70.25\pm0.78$ km/s/Mpc at 68\% CL, which is in tension with R19 at $2.3$ standard deviation, much lower than the present tension with $\Lambda$CDM and many other dark energy models, suggesting a formidable alleviation of the Hubble tension with phantom crossing. Finally, in Table~\ref{table_chi} we can see that the Phantom Crossing fits better than $\Lambda$CDM the full dataset combination, improving the $\chi^2_{bf}$.

\acknowledgements{EDV acknowledges the support of the Addison-Wheeler Fellowship awarded by the Institute of Advanced Study at Durham University.
AM acknowledges the financial support from the Science and Engineering Research Board (SERB), Department of Science and Technology,
Government of India as a National Post-Doctoral Fellow (NPDF, File no. PDF/2018/001859). AAS acknowledges funding from DST-SERB, Govt of India, under the project NO. MTR/20l9/000599. The authors also thank Ruchika for computational help.}


\begin{thebibliography}{999}

%\cite{Riess:1998cb}
\bibitem{Riess:1998cb} 
  A.~G.~Riess {\it et al.} [Supernova Search Team],
  %``Observational evidence from supernovae for an accelerating universe and a cosmological constant,''
  Astron.\ J.\  {\bf 116}, 1009 (1998).
  %doi:10.1086/300499
  %[astro-ph/9805201].
  %%CITATION = doi:10.1086/300499;%%
  %12140 citations counted in INSPIRE as of 19 Feb 2020
  
  %\cite{Perlmutter:1998np}
\bibitem{Perlmutter:1998np} 
  S.~Perlmutter {\it et al.} [Supernova Cosmology Project Collaboration],
  %``Measurements of $\Omega$ and $\Lambda$ from 42 high redshift supernovae,''
  Astrophys.\ J.\  {\bf 517}, 565 (1999).
  %doi:10.1086/307221
  %[astro-ph/9812133].
  %%CITATION = doi:10.1086/307221;%%
  %12272 citations counted in INSPIRE as of 19 Feb 2020
  
  
  
  
  %\cite{Bamba:2012cp}
\bibitem{Bamba:2012cp}
K.~Bamba, S.~Capozziello, S.~Nojiri and S.~D.~Odintsov,
%``Dark energy cosmology: the equivalent description via different theoretical models and cosmography tests,''
Astrophys. Space Sci. \textbf{342}, 155 (2012).
%doi:10.1007/s10509-012-1181-8
%[arXiv:1205.3421 [gr-qc]].


%\cite{Carroll:2000fy}
\bibitem{Carroll:2000fy} 
  S.~M.~Carroll,
  %``The Cosmological constant,''
  Living Rev.\ Rel.\  {\bf 4}, 1 (2001).
  %doi:10.12942/lrr-2001-1
  %[astro-ph/0004075].
  %%CITATION = doi:10.12942/lrr-2001-1;%%
  %1508 citations counted in INSPIRE as of 19 Feb 2020



%\cite{Peebles:2002gy}
\bibitem{Peebles:2002gy} 
  P.~J.~E.~Peebles and B.~Ratra,
  %``The Cosmological Constant and Dark Energy,''
  Rev.\ Mod.\ Phys.\  {\bf 75}, 559 (2003).
  %doi:10.1103/RevModPhys.75.559
  %[astro-ph/0207347].
  %%CITATION = doi:10.1103/RevModPhys.75.559;%%
  %3546 citations counted in INSPIRE as of 19 Feb 2020

%\cite{Padmanabhan:2002ji}
\bibitem{Padmanabhan:2002ji} 
  T.~Padmanabhan,
  %``Cosmological constant: The Weight of the vacuum,''
  Phys.\ Rept.\  {\bf 380}, 235 (2003).
  %doi:10.1016/S0370-1573(03)00120-0
  %[hep-th/0212290].
  %%CITATION = doi:10.1016/S0370-1573(03)00120-0;%%
  %2503 citations counted in INSPIRE as of 19 Feb 2020
  
  
  %\cite{Copeland:2006wr}
\bibitem{Copeland:2006wr} 
  E.~J.~Copeland, M.~Sami and S.~Tsujikawa,
  %``Dynamics of dark energy,''
  Int.\ J.\ Mod.\ Phys.\ D {\bf 15}, 1753 (2006).
  %doi:10.1142/S021827180600942X
  %[hep-th/0603057].
  %%CITATION = doi:10.1142/S021827180600942X;%%
  %4081 citations counted in INSPIRE as of 19 Feb 2020
  
  
  
%\cite{Kamenshchik:2001cp}
\bibitem{Kamenshchik:2001cp} 
  A.~Y.~Kamenshchik, U.~Moschella and V.~Pasquier,
  %``An Alternative to quintessence,''
  Phys.\ Lett.\ B {\bf 511}, 265 (2001).
  %doi:10.1016/S0370-2693(01)00571-8
  %[gr-qc/0103004].
  %%CITATION = doi:10.1016/S0370-2693(01)00571-8;%%
  %1767 citations counted in INSPIRE as of 19 Feb 2020
  
%\cite{Bento:2002ps}
\bibitem{Bento:2002ps} 
  M.~C.~Bento, O.~Bertolami and A.~A.~Sen,
  %``Generalized Chaplygin gas, accelerated expansion and dark energy matter unification,''
  Phys.\ Rev.\ D {\bf 66}, 043507 (2002).
  %doi:10.1103/PhysRevD.66.043507
  %[gr-qc/0202064].
  %%CITATION = doi:10.1103/PhysRevD.66.043507;%%
  %1397 citations counted in INSPIRE as of 19 Feb 2020  
  
  %\cite{Akrami:2018vks}
\bibitem{Akrami:2018vks}
Y.~Akrami \textit{et al.} [Planck],
%``Planck 2018 results. I. Overview and the cosmological legacy of Planck,''
[arXiv:1807.06205 [astro-ph.CO]].
  
  
\bibitem{SuzukiSCP}  
 N. Suzuki {\it et al.},
 % "The Hubble space telescope cluster Supernova survey: V. Improving the dark energy constraints
% above z&gt;1 and building an early-type-hosted Supernova sample",
Astrophys. J. {\bf 746}, 85 (2012). 


\bibitem{DawsonBOSS}
 K. S. Dawson {\it et al.},
 % "The Baryon Oscillation Spectroscopic Survey of SDSS-III"
Astron. J. {\bf 145}, 10 (2012). 
%arXiv:1208.0022

\bibitem{AlamSDSS}
S. Alam et al. (SDSS-III Collaboration), 
%The eleventh and twelfth data releases of the sloan digital sky survey: Final data from SDSS-III,
Astrophys. J. Suppl. Ser. {\bf 219}, 12 (2015).
%arXiv:1501.00963

%\cite{Vikman:2004dc}
\bibitem{Vikman:2004dc}
A.~Vikman,
%``Can dark energy evolve to the phantom?,''
Phys. Rev. D \textbf{71}, 023515 (2005).
%doi:10.1103/PhysRevD.71.023515
%[arXiv:astro-ph/0407107 [astro-ph]].
%467 citations counted in INSPIRE as of 09 Sep 2020

%\cite{Nojiri:2005sr}
\bibitem{Nojiri:2005sr}
S.~Nojiri and S.~D.~Odintsov,
%``Inhomogeneous equation of state of the universe: Phantom era, future singularity and crossing the phantom barrier,''
Phys. Rev. D \textbf{72}, 023003 (2005).
%doi:10.1103/PhysRevD.72.023003
%[arXiv:hep-th/0505215 [hep-th]].

%\cite{Nojiri:2005pu}
\bibitem{Nojiri:2005pu}
S.~Nojiri and S.~D.~Odintsov,
%``Unifying phantom inflation with late-time acceleration: Scalar phantom-non-phantom transition model and generalized holographic dark energy,''
Gen. Rel. Grav. \textbf{38}, 1285 (2006).
%doi:10.1007/s10714-006-0301-6
%[arXiv:hep-th/0506212 [hep-th]].

%\cite{Nojiri:2006ww}
\bibitem{Nojiri:2006ww}
S.~Nojiri and S.~D.~Odintsov,
%``The Oscillating dark energy: Future singularity and coincidence problem,''
Phys. Lett. B \textbf{637}, 139 (2006).
%doi:10.1016/j.physletb.2006.04.026
%[arXiv:hep-th/0603062 [hep-th]].

%\cite{Bamba:2008hq}
\bibitem{Bamba:2008hq}
K.~Bamba, C.~Q.~Geng, S.~Nojiri and S.~D.~Odintsov,
%``Crossing of the phantom divide in modified gravity,''
Phys. Rev. D \textbf{79}, 083014 (2009).
%doi:10.1103/PhysRevD.79.083014
%[arXiv:0810.4296 [hep-th]].

\bibitem{jaime}
Luisa~G.~Jaime, M.~Jaber, and Celia~Escamilla-Rivera,
Phys. Rev. D \textbf{98}, 083530 (2018).

\bibitem{sarikd}
Emmanuel N. Saridakis,
Class. Quant. Grav. \textbf{30}, 075003 (2013).

%\cite{Chimento:2008ws}
\bibitem{Chimento:2008ws}
L.~P.~Chimento, M.~I.~Forte, R.~Lazkoz and M.~G.~Richarte,
%``Internal space structure generalization of the quintom cosmological scenario,''
Phys. Rev. D \textbf{79}, 043502 (2009).
%doi:10.1103/PhysRevD.79.043502
%[arXiv:0811.3643 [astro-ph]].

\bibitem{Hu:2004kh}
W.~Hu,
%``Crossing the phantom divide: Dark energy internal degrees of freedom,''
Phys. Rev. D \textbf{71}, 047301 (2005).
%doi:10.1103/PhysRevD.71.047301
%[arXiv:astro-ph/0410680 [astro-ph]].



\bibitem{Wei:2005si}
H.~Wei and R.~G.~Cai,
%``A note on crossing the phantom divide in hybrid dark energy model,''
Phys. Lett. B \textbf{634}, 9 (2006).
%doi:10.1016/j.physletb.2006.01.043
%[arXiv:astro-ph/0512018 [astro-ph]].

%\cite{Deffayet:2010qz}
\bibitem{Deffayet:2010qz}
C.~Deffayet, O.~Pujolas, I.~Sawicki and A.~Vikman,
%``Imperfect Dark Energy from Kinetic Gravity Braiding,''
JCAP \textbf{10}, 026 (2010).
%doi:10.1088/1475-7516/2010/10/026
%[arXiv:1008.0048 [hep-th]].
%505 citations counted in INSPIRE as of 09 Sep 2020


\bibitem{Matsumoto:2017qil}
J.~Matsumoto,
%``Phantom crossing dark energy in Horndeski’s theory,''
Phys. Rev. D \textbf{97}, 123538 (2018).
%doi:10.1103/PhysRevD.97.123538
%[arXiv:1712.10015 [gr-qc]].


\bibitem{Nesseris:2006er}
S.~Nesseris and L.~Perivolaropoulos,
%``Crossing the Phantom Divide: Theoretical Implications and Observational Status,''
JCAP \textbf{01}, 018 (2007).
%doi:10.1088/1475-7516/2007/01/018
%[arXiv:astro-ph/0610092 [astro-ph]].

\bibitem{Fang:2008sn}
W.~Fang, W.~Hu and A.~Lewis,
%``Crossing the Phantom Divide with Parameterized Post-Friedmann Dark Energy,''
Phys. Rev. D \textbf{78}, 087303 (2008).
%doi:10.1103/PhysRevD.78.087303
%[arXiv:0808.3125 [astro-ph]].

\bibitem{Shafer:2013pxa}
D.~L.~Shafer and D.~Huterer,
%``Chasing the phantom: A closer look at Type Ia supernovae and the dark energy equation of state,''
Phys. Rev. D \textbf{89}, 063510 (2014).
%doi:10.1103/PhysRevD.89.063510
%[arXiv:1312.1688 [astro-ph.CO]].


\bibitem{Wang:2018fng}
Y.~Wang, L.~Pogosian, G.~B.~Zhao and A.~Zucca,
%``Evolution of dark energy reconstructed from the latest observations,''
Astrophys. J. Lett. \textbf{869}, L8 (2018).
%doi:10.3847/2041-8213/aaf238
%[arXiv:1807.03772 [astro-ph.CO]].



\bibitem{zhaoPRL}G.-B. Zhao, R. G. Crittenden, L. Pogosian and X. Zhang,
%Examining the Evidence for Dynamical Dark Energy. 
Phys. Rev. Lett. {\bf 109}, 171301 (2012).
%[arXiv:1207.3804 [astro-ph.CO]]




\bibitem{zhao}G.-B.  Zhao et al. Nat.\ Astron. {\bf 1}, 627 (2017).


%\cite{Capozziello:2018jya}
\bibitem{Capozziello:2018jya}
S.~Capozziello, Ruchika and A.~A.~Sen,
%``Model independent constraints on dark energy evolution from low-redshift observations,''
Mon. Not. Roy. Astron. Soc. \textbf{484} 4484 (2019).
%doi:10.1093/mnras/stz176
%[arXiv:1806.03943 [astro-ph.CO]].


  %\cite{Riess:2019cxk}
\bibitem{Riess:2019cxk} 
  A.~G.~Riess, S.~Casertano, W.~Yuan, L.~M.~Macri and D.~Scolnic,
  %``Large Magellanic Cloud Cepheid Standards Provide a 1% Foundation for the Determination of the Hubble Constant and Stronger Evidence for Physics beyond $\Lambda$CDM,''
  Astrophys.\ J.\  {\bf 876}, 85 (2019).
  %doi:10.3847/1538-4357/ab1422
  %[arXiv:1903.07603 [astro-ph.CO]].
  %%CITATION = doi:10.3847/1538-4357/ab1422;%%
  
    %\cite{Aghanim:2018eyx}
\bibitem{Aghanim:2018eyx} 
  N.~Aghanim {\it et al.} [Planck Collaboration],
  %``Planck 2018 results. VI. Cosmological parameters,''
  arXiv:1807.06209 [astro-ph.CO].
  %%CITATION = ARXIV:1807.06209;%%
  %2298 citations counted in INSPIRE as of 19 Feb 2020


\bibitem{DiValentino:2016hlg} 
  E.~Di Valentino, A.~Melchiorri and J.~Silk,
  %{\it Reconciling Planck with the local value of $H_0$ in extended parameter space,}
  Phys.\ Lett.\ B {\bf 761}, 242 (2016).
  %doi:10.1016/j.physletb.2016.08.043
  %[arXiv:1606.00634 [astro-ph.CO]].
  
  
  %\cite{Bernal:2016gxb}
\bibitem{Bernal:2016gxb} 
  J.~L.~Bernal, L.~Verde and A.~G.~Riess,
  %``The trouble with $H_0$,''
  JCAP {\bf 1610}, 019 (2016).
  %doi:10.1088/1475-7516/2016/10/019
  %[arXiv:1607.05617 [astro-ph.CO]].
  %%CITATION = doi:10.1088/1475-7516/2016/10/019;%%
  
  %\cite{Kumar:2016zpg}
\bibitem{Kumar:2016zpg} 
  S.~Kumar and R.~C.~Nunes,
  %``Probing the interaction between dark matter and dark energy in the presence of massive neutrinos,''
  Phys.\ Rev.\ D {\bf 94}, 123511 (2016).
  %doi:10.1103/PhysRevD.94.123511
  %[arXiv:1608.02454 [astro-ph.CO]].
  %%CITATION = doi:10.1103/PhysRevD.94.123511;%%
  
  \bibitem{Kumar:2017dnp} 
  S.~Kumar and R.~C.~Nunes,
  %``Echo of interactions in the dark sector,''
  Phys.\ Rev.\ D {\bf 96}, 103511 (2017).
  %doi:10.1103/PhysRevD.96.103511
  %[arXiv:1702.02143 [astro-ph.CO]]. 
  
  \bibitem{DiValentino:2017iww} 
  E.~Di Valentino, A.~Melchiorri and O.~Mena,
  %``Can interacting dark energy solve the $H_0$ tension?,''
  Phys.\ Rev.\ D {\bf 96}, 043503 (2017).
  %doi:10.1103/PhysRevD.96.043503
  %[arXiv:1704.08342 [astro-ph.CO]].

%\cite{DiValentino:2017oaw}
\bibitem{DiValentino:2017oaw} 
  E.~Di Valentino, C.~Bøehm, E.~Hivon and F.~R.~Bouchet,
  %``Reducing the $H_0$ and $\sigma_8$ tensions with Dark Matter-neutrino interactions,''
  Phys.\ Rev.\ D {\bf 97}, 043513 (2018).
  %doi:10.1103/PhysRevD.97.043513
  %[arXiv:1710.02559 [astro-ph.CO]].
  %%CITATION = doi:10.1103/PhysRevD.97.043513;%%
  
  \bibitem{DiValentino:2017rcr} 
  E.~Di Valentino, E.~V.~Linder and A.~Melchiorri,
  %``Vacuum phase transition solves the $H_0$ tension,''
  Phys.\ Rev.\ D {\bf 97}, 043528 (2018).
  %doi:10.1103/PhysRevD.97.043528
  %[arXiv:1710.02153 [astro-ph.CO]].

%\cite{DiValentino:2017zyq}
\bibitem{DiValentino:2017zyq} 
  E.~Di Valentino, A.~Melchiorri, E.~V.~Linder and J.~Silk,
  %``Constraining Dark Energy Dynamics in Extended Parameter Space,''
  Phys.\ Rev.\ D {\bf 96}, 023523 (2017).
  %doi:10.1103/PhysRevD.96.023523
  %[arXiv:1704.00762 [astro-ph.CO]].
  %%CITATION = doi:10.1103/PhysRevD.96.023523;%%

  
  %\cite{Sola:2017znb}
\bibitem{Sola:2017znb} 
  J.~Solà, A.~Gómez-Valent and J.~de Cruz Pérez,
  %``The $H_0$ tension in light of vacuum dynamics in the Universe,''
  Phys.\ Lett.\ B {\bf 774}, 317 (2017).
  %doi:10.1016/j.physletb.2017.09.073
  %[arXiv:1705.06723 [astro-ph.CO]].
  %%CITATION = doi:10.1016/j.physletb.2017.09.073;%%

%\cite{Nunes:2018xbm}
\bibitem{Nunes:2018xbm} 
  R.~C.~Nunes,
  %``Structure formation in $f(T)$ gravity and a solution for $H_0$ tension,''
  JCAP {\bf 1805}, 052 (2018).
  %doi:10.1088/1475-7516/2018/05/052
  %[arXiv:1802.02281 [gr-qc]].
  %%CITATION = doi:10.1088/1475-7516/2018/05/052;%%

%\cite{Yang:2018euj}
\bibitem{Yang:2018euj} 
  W.~Yang, S.~Pan, E.~Di Valentino, R.~C.~Nunes, S.~Vagnozzi and D.~F.~Mota,
  %``Tale of stable interacting dark energy, observational signatures, and the $H_0$ tension,''
  JCAP {\bf 1809}, 019 (2018).
  %doi:10.1088/1475-7516/2018/09/019
  %[arXiv:1805.08252 [astro-ph.CO]].
  %%CITATION = doi:10.1088/1475-7516/2018/09/019;%%
  
  \bibitem{Yang:2018uae}
  W.~Yang, A.~Mukherjee, E.~Di Valentino and S.~Pan,
 %{\it Interacting dark energy with time varying equation of state and the $H_0$ tension,}
  Phys.\ Rev.\ D {\bf 98},  123527 (2018).
  %doi:10.1103/PhysRevD.98.123527
  %[arXiv:1809.06883 [astro-ph.CO]].
  
  \bibitem{Yang:2018qmz} 
  W.~Yang, S.~Pan, E.~Di Valentino, E.~N.~Saridakis and S.~Chakraborty,
  %{\it Observational constraints on one-parameter dynamical dark-energy parametrizations and the $H_0$ tension,}
  Phys.\ Rev.\ D {\bf 99}, 043543 (2019).
  %doi:10.1103/PhysRevD.99.043543
  %[arXiv:1810.05141 [astro-ph.CO]].  


   \bibitem{Poulin:2018cxd} 
  V.~Poulin, T.~L.~Smith, T.~Karwal and M.~Kamionkowski,
  %{\it Early Dark Energy Can Resolve The Hubble Tension,}
  Phys.\ Rev.\ Lett.\  {\bf 122}, 221301 (2019).
  %doi:10.1103/PhysRevLett.122.221301
  %[arXiv:1811.04083 [astro-ph.CO]].
  
 \bibitem{Mortsell:2018mfj} 
  E.~M\"{o}rtsell and S.~Dhawan,
  %{\it Does the Hubble constant tension call for new physics?,}
  JCAP {\bf 1809}, 025 (2018).
  %doi:10.1088/1475-7516/2018/09/025
  %[arXiv:1801.07260 [astro-ph.CO]].

%\cite{Martinelli:2019dau}
\bibitem{Martinelli:2019dau} 
  M.~Martinelli, N.~B.~Hogg, S.~Peirone, M.~Bruni and D.~Wands,
  %``Constraints on the interacting vacuum–geodesic CDM scenario,''
  Mon.\ Not.\ Roy.\ Astron.\ Soc.\  {\bf 488}, 3423 (2019).
  %doi:10.1093/mnras/stz1915
  %[arXiv:1902.10694 [astro-ph.CO]].
  %%CITATION = doi:10.1093/mnras/stz1915;%%

\bibitem{Vattis:2019efj} 
  K.~Vattis, S.~M.~Koushiappas and A.~Loeb,
  %``Dark matter decaying in the late Universe can relieve the H0 tension,''
  Phys.\ Rev.\ D {\bf 99}, 121302 (2019).
  %doi:10.1103/PhysRevD.99.121302
  %[arXiv:1903.06220 [astro-ph.CO]].
  
  \bibitem{Kumar:2019wfs} 
  S.~Kumar, R.~C.~Nunes and S.~K.~Yadav,
  %``Dark sector interaction: a remedy of the tensions between CMB and LSS data,''
  Eur.\ Phys.\ J.\ C {\bf 79}, 576 (2019).
  %doi:10.1140/epjc/s10052-019-7087-7
  %[arXiv:1903.04865 [astro-ph.CO]].
  
  %\cite{Agrawal:2019lmo}
\bibitem{Agrawal:2019lmo} 
  P.~Agrawal, F.~Y.~Cyr-Racine, D.~Pinner and L.~Randall,
  %``Rock 'n' Roll Solutions to the Hubble Tension,''
  arXiv:1904.01016 [astro-ph.CO].
  %%CITATION = ARXIV:1904.01016;%%
  
  %\cite{Yang:2019qza}
\bibitem{Yang:2019qza} 
  W.~Yang, S.~Pan, E.~Di Valentino, A.~Paliathanasis and J.~Lu,
  %``Challenging bulk viscous unified scenarios with cosmological observations,''
  Phys.\ Rev.\ D {\bf 100}, 103518 (2019).
  %doi:10.1103/PhysRevD.100.103518
  %[arXiv:1906.04162 [astro-ph.CO]].
  %%CITATION = doi:10.1103/PhysRevD.100.103518;%%
  
  \bibitem{Yang:2019uzo} 
  W.~Yang, O.~Mena, S.~Pan and E.~Di Valentino,
  %``Dark sectors with dynamical coupling,''
  Phys.\ Rev.\ D {\bf 100}, 083509 (2019).
  %doi:10.1103/PhysRevD.100.083509
  %[arXiv:1906.11697 [astro-ph.CO]].
  
  %\cite{DiValentino:2019exe}
\bibitem{DiValentino:2019exe} 
  E.~Di Valentino, R.~Z.~Ferreira, L.~Visinelli and U.~Danielsson,
  %``Late time transitions in the quintessence field and the $H_0$ tension,''
  Phys.\ Dark Univ.\  {\bf 26}, 100385 (2019).
  %doi:10.1016/j.dark.2019.100385
  %[arXiv:1906.11255 [astro-ph.CO]].
  %%CITATION = doi:10.1016/j.dark.2019.100385;%%
  
  \bibitem{Pan:2019gop} 
  S.~Pan, W.~Yang, E.~Di Valentino, E.~N.~Saridakis and S.~Chakraborty,
 %{\it Interacting scenarios with dynamical dark energy: observational constraints and alleviation of the $H_0$ tension,}
 Phys. Rev. D \textbf{100}, 103520 (2019)
 %doi:10.1103/PhysRevD.100.103520
 % arXiv:1907.07540 [astro-ph.CO]. 
  
  %\cite{Martinelli:2019krf}
\bibitem{Martinelli:2019krf} 
  M.~Martinelli and I.~Tutusaus,
  %``CMB tensions with low-redshift $H_0$ and $S_8$ measurements: impact of a redshift-dependent type-Ia supernovae intrinsic luminosity,''
  Symmetry {\bf 11}, 986 (2019).
  %doi:10.3390/sym11080986
  %[arXiv:1906.09189 [astro-ph.CO]].
  %%CITATION = doi:10.3390/sym11080986;%%
  
  
  
    \bibitem{Pan:2019hac} 
  S.~Pan, W.~Yang, E.~Di Valentino, A.~Shafieloo and S.~Chakraborty,
  %{\it Reconciling $H_0$ tension in a six parameter space?,}
  JCAP \textbf{06}, 062 (2020).
 %doi:10.1088/1475-7516/2020/06/062
 % arXiv:1907.12551 [astro-ph.CO].
  
  %\cite{DiValentino:2019dzu}
\bibitem{DiValentino:2019dzu} 
  E.~Di Valentino, A.~Melchiorri and J.~Silk,
  %``Cosmological constraints in extended parameter space from the Planck 2018 Legacy release,''
  JCAP {\bf 2001}, 013 (2020).
  %doi:10.1088/1475-7516/2020/01/013
  %[arXiv:1908.01391 [astro-ph.CO]].
  %%CITATION = doi:10.1088/1475-7516/2020/01/013;%%
  
  
  
  %\cite{DiValentino:2019ffd}
\bibitem{DiValentino:2019ffd} 
  E.~Di Valentino, A.~Melchiorri, O.~Mena and S.~Vagnozzi,
  %``Interacting dark energy after the latest Planck, DES, and $H_0$ measurements: an excellent solution to the $H_0$ and cosmic shear tensions,''
  Phys. Dark Univ. \textbf{30}, 100666 (2020).
  % doi:10.1016/j.dark.2020.100666
  % arXiv:1908.04281 [astro-ph.CO].
  %%CITATION = ARXIV:1908.04281;%%
  
  %\cite{DiValentino:2019jae}
\bibitem{DiValentino:2019jae} 
  E.~Di Valentino, A.~Melchiorri, O.~Mena and S.~Vagnozzi,
  %``Non-minimal dark sector physics and cosmological tensions,''
  Phys.\ Rev.\ D {\bf 101}, 063502 (2020).
  %doi:10.1103/PhysRevD.101.063502
  %[arXiv:1910.09853 [astro-ph.CO]].
  %%CITATION = doi:10.1103/PhysRevD.101.063502;%%
  
  %\cite{Colgain:2019joh}
\bibitem{Colgain:2019joh} 
  E.~Ó.~Colgáin and H.~Yavartanoo,
  %``Testing the Swampland: $H_0$ tension,''
  Phys.\ Lett.\ B {\bf 797}, 134907 (2019).
  %doi:10.1016/j.physletb.2019.134907
  %[arXiv:1905.02555 [astro-ph.CO]].
  %%CITATION = doi:10.1016/j.physletb.2019.134907;%%
  
  %\cite{Alcaniz:2019kah}
\bibitem{Alcaniz:2019kah}
J.~Alcaniz, N.~Bernal, A.~Masiero and F.~S.~Queiroz,
%``Light Dark Matter: A Common Solution to the Lithium and ${H_0}$ Problems,''
[arXiv:1912.05563 [astro-ph.CO]].
%10 citations counted in INSPIRE as of 09 Sep 2020
  
  %\cite{Pan:2019jqh}
\bibitem{Pan:2019jqh} 
  S.~Pan, W.~Yang, C.~Singha and E.~N.~Saridakis,
  %``Observational constraints on sign-changeable interaction models and alleviation of the $H_0$ tension,''
  Phys.\ Rev.\ D {\bf 100}, 083539 (2019).
  %doi:10.1103/PhysRevD.100.083539
  %[arXiv:1903.10969 [astro-ph.CO]].
  %%CITATION = doi:10.1103/PhysRevD.100.083539;%%

%\cite{Berghaus:2019cls}
\bibitem{Berghaus:2019cls} 
  K.~V.~Berghaus and T.~Karwal,
  %``Thermal Friction as a Solution to the Hubble Tension,''
  Phys. Rev. D \textbf{101}, 083537 (2020)
  % doi:10.1103/PhysRevD.101.083537
  % arXiv:1911.06281 [astro-ph.CO].
  %%CITATION = ARXIV:1911.06281;%%

%\cite{Knox:2019rjx}
\bibitem{Knox:2019rjx} 
  L.~Knox and M.~Millea,
  %``Hubble constant hunter’s guide,''
  Phys.\ Rev.\ D {\bf 101}, 043533 (2020).
  %doi:10.1103/PhysRevD.101.043533
  %[arXiv:1908.03663 [astro-ph.CO]].
  %%CITATION = doi:10.1103/PhysRevD.101.043533;%%
  
  %\cite{Pandey:2019plg}
\bibitem{Pandey:2019plg} 
  K.~L.~Pandey, T.~Karwal and S.~Das,
  %``Alleviating the $H_0$ and $\sigma_8$ anomalies with a decaying dark matter model,''
  JCAP \textbf{07}, 026 (2020).
  % doi:10.1088/1475-7516/2020/07/026
  % arXiv:1902.10636 [astro-ph.CO].
  %%CITATION = ARXIV:1902.10636;%%

%\cite{Adhikari:2019fvb}
\bibitem{Adhikari:2019fvb} 
  S.~Adhikari and D.~Huterer,
  %``Super-CMB fluctuations can resolve the Hubble tension,''
  Phys. Dark Univ. \textbf{28}, 100539 (2020).
  % doi:10.1016/j.dark.2020.100539
  % arXiv:1905.02278 [astro-ph.CO].
  %%CITATION = ARXIV:1905.02278;%%
  
  %\cite{Hart:2019dxi}
\bibitem{Hart:2019dxi} 
  L.~Hart and J.~Chluba,
  %``Updated fundamental constant constraints from Planck 2018 data and possible relations to the Hubble tension,''
  %doi:10.1093/mnras/staa412
  Mon. Not. Roy. Astron. Soc. \textbf{493}, 3255 (2020).
 % doi:10.1093/mnras/staa412
 % arXiv:1912.03986 [astro-ph.CO].

%\cite{Liao:2020zko}
\bibitem{Liao:2020zko} 
  K.~Liao, A.~Shafieloo, R.~E.~Keeley and E.~V.~Linder,
  %``Determining $H_0$ Model-Independently and Consistency Tests,''
  arXiv:2002.10605 [astro-ph.CO].
  
  %\cite{Benevento:2020fev}
\bibitem{Benevento:2020fev} 
  G.~Benevento, W.~Hu and M.~Raveri,
  %``Can Late Dark Energy Transitions Raise the Hubble constant?,''
  Astrophys. J. Lett. \textbf{895}, L29 (2020).
  % doi:10.3847/2041-8213/ab8dbb 
 %  arXiv:2002.11707 [astro-ph.CO].
  
  \bibitem{Vagnozzi:2019ezj} 
  S.~Vagnozzi,
  %{\it New physics in light of the $H_0$ tension: an alternative view,}
  Phys. Rev. D \textbf{102}, 023518 (2020).
 %doi:10.1103/PhysRevD.102.023518
 % arXiv:1907.07569 [astro-ph.CO].]
  
  %\cite{Chudaykin:2020acu}
\bibitem{Chudaykin:2020acu}
A.~Chudaykin, D.~Gorbunov and N.~Nedelko,
%``Combined analysis of Planck and SPTPol data favors the early dark energy models,''
JCAP \textbf{08}, 013 (2020).
%doi:10.1088/1475-7516/2020/08/013
%[arXiv:2004.13046 [astro-ph.CO]].

%\cite{Alestas:2020mvb}
\bibitem{Alestas:2020mvb}
G.~Alestas, L.~Kazantzidis and L.~Perivolaropoulos,
%``$H_0$ Tension, Phantom Dark Energy and Cosmological Parameter Degeneracies,''
Phys. Rev. D \textbf{101}, 123516 (2020).
%doi:10.1103/PhysRevD.101.123516
%[arXiv:2004.08363 [astro-ph.CO]].
  
  %\cite{Wang:2020zfv}
\bibitem{Wang:2020zfv} 
  D.~Wang and D.~Mota,
  %``Can $f(T)$ gravity resolve the $H_0$ tension?,''
  Phys. Rev. D \textbf{102}, 063530 (2020).
  % doi:10.1103/PhysRevD.102.063530
  % arXiv:2003.10095 [astro-ph.CO].
  %%CITATION = ARXIV:2003.10095;%%
  
    %\cite{Blinov:2020hmc}
\bibitem{Blinov:2020hmc} 
  N.~Blinov and G.~Marques-Tavares,
  %``Interacting radiation after Planck and its implications for the Hubble Tension,''
  JCAP \textbf{09}, 029 (2020).
  % doi:10.1088/1475-7516/2020/09/029
  %arXiv:2003.08387 [astro-ph.CO].
  %%CITATION = ARXIV:2003.08387;%%
  
  %\cite{DiValentino:2020zio}
\bibitem{DiValentino:2020zio}
E.~Di Valentino, L.~A.~Anchordoqui, O.~Akarsu, Y.~Ali-Haimoud, L.~Amendola, N.~Arendse, M.~Asgari, M.~Ballardini, S.~Basilakos and E.~Battistelli, \textit{et al.}
%``Cosmology Intertwined II: The Hubble Constant Tension,''
[arXiv:2008.11284 [astro-ph.CO]].

%\cite{DiValentino:2021izs}
\bibitem{DiValentino:2021izs}
E.~Di Valentino, O.~Mena, S.~Pan, L.~Visinelli, W.~Yang, A.~Melchiorri, D.~F.~Mota, A.~G.~Riess and J.~Silk,
%``In the Realm of the Hubble tension $-$ a Review of Solutions,''
[arXiv:2103.01183 [astro-ph.CO]].
  
\bibitem{Alestas:2020zol}
G.~Alestas, L.~Kazantzidis, L.~Perivolaropoulos,
%    title = "{A $w-M$ phantom transition at $z_t<0.1$ as a resolution of the Hubble tension}",
arXiv:2012.13932 [astro-ph.CO].
    
\bibitem{Camarena:2021jlr}
D.~Camarena,and V.~Marra,
    %title = "{On the use of the local prior on the absolute magnitude of Type Ia supernovae in cosmological inference}",
 arXiv:2101.08641[astro-ph.CO].  
 
\bibitem{sen_pressure}
A.~A.~Sen,
Phys. Rev. D \textbf{77}, 043508, (2008).
  
  %\cite{Delubac:2014aqe}
\bibitem{Delubac:2014aqe}
T.~Delubac \textit{et al.} [BOSS],
%``Baryon acoustic oscillations in the Lyα forest of BOSS DR11 quasars,''
Astron. Astrophys. \textbf{574}, A59 (2015).
%doi:10.1051/0004-6361/201423969
%[arXiv:1404.1801 [astro-ph.CO]].
  
  %\cite{Poulin:2018zxs}
\bibitem{Poulin:2018zxs}
V.~Poulin, K.~K.~Boddy, S.~Bird and M.~Kamionkowski,
%``Implications of an extended dark energy cosmology with massive neutrinos for cosmological tensions,''
Phys. Rev. D \textbf{97}, 123504 (2018).
%doi:10.1103/PhysRevD.97.123504
%[arXiv:1803.02474 [astro-ph.CO]].



%\cite{Dutta:2018vmq}
\bibitem{Dutta:2018vmq}
K.~Dutta, Ruchika, A.~Roy, A.~A.~Sen and M.~M.~Sheikh-Jabbari,
%``Beyond $\Lambda $CDM with low and high redshift data: implications for dark energy,''
Gen. Rel. Grav. \textbf{52}, 15 (2020).
%doi:10.1007/s10714-020-2665-4
%[arXiv:1808.06623 [astro-ph.CO]].


    
  
%\cite{Aghanim:2019ame}
\bibitem{Aghanim:2019ame} 
  N.~Aghanim {\it et al.} [Planck Collaboration],
  %``Planck 2018 results. V. CMB power spectra and likelihoods,''
  arXiv:1907.12875 [astro-ph.CO].
  %%CITATION = ARXIV:1907.12875;%%
  
%\cite{Efstathiou:2019mdh}
\bibitem{Efstathiou:2019mdh}
G.~Efstathiou and S.~Gratton,
%``A Detailed Description of the CamSpec Likelihood Pipeline and a Reanalysis of the Planck High Frequency Maps,''
[arXiv:1910.00483 [astro-ph.CO]].

  
  %\cite{Beutler:2011hx}
\bibitem{Beutler:2011hx} 
  F.~Beutler {\it et al.},
  %``The 6dF Galaxy Survey: Baryon Acoustic Oscillations and the Local Hubble Constant,''
  Mon.\ Not.\ Roy.\ Astron.\ Soc.\  {\bf 416}, 3017 (2011).
  %doi:10.1111/j.1365-2966.2011.19250.x
  %[arXiv:1106.3366 [astro-ph.CO]].
  %%CITATION = doi:10.1111/j.1365-2966.2011.19250.x;%%
  
  %\cite{Ross:2014qpa}
\bibitem{Ross:2014qpa} 
  A.~J.~Ross, L.~Samushia, C.~Howlett, W.~J.~Percival, A.~Burden and M.~Manera,
  %``The clustering of the SDSS DR7 main Galaxy sample – I. A 4 per cent distance measure at $z = 0.15$,''
  Mon.\ Not.\ Roy.\ Astron.\ Soc.\  {\bf 449}, 835 (2015).
  %doi:10.1093/mnras/stv154
  %[arXiv:1409.3242 [astro-ph.CO]].
  %%CITATION = doi:10.1093/mnras/stv154;%%
  
  %\cite{Alam:2016hwk}
\bibitem{Alam:2016hwk} 
  S.~Alam {\it et al.} [BOSS Collaboration],
  %``The clustering of galaxies in the completed SDSS-III Baryon Oscillation Spectroscopic Survey: cosmological analysis of the DR12 galaxy sample,''
  Mon.\ Not.\ Roy.\ Astron.\ Soc.\  {\bf 470}, 2617 (2017).
  %doi:10.1093/mnras/stx721
  %[arXiv:1607.03155 [astro-ph.CO]].
  %%CITATION = doi:10.1093/mnras/stx721;%%
  
%\cite{deMattia:2020fkb}
\bibitem{deMattia:2020fkb}
A.~de Mattia, V.~Ruhlmann-Kleider, A.~Raichoor, A.~J.~Ross, A.~Tamone, C.~Zhao, S.~Alam, S.~Avila, E.~Burtin and J.~Bautista, \textit{et al.}
%``The Completed SDSS-IV extended Baryon Oscillation Spectroscopic Survey: measurement of the BAO and growth rate of structure of the emission line galaxy sample from the anisotropic power spectrum between redshift 0.6 and 1.1,''
doi:10.1093/mnras/staa3891
[arXiv:2007.09008 [astro-ph.CO]].
  
  %\cite{Scolnic:2017caz}
\bibitem{Scolnic:2017caz} 
  D.~M.~Scolnic {\it et al.},
  %``The Complete Light-curve Sample of Spectroscopically Confirmed SNe Ia from Pan-STARRS1 and Cosmological Constraints from the Combined Pantheon Sample,''
  Astrophys.\ J.\  {\bf 859}, 101 (2018).
  %doi:10.3847/1538-4357/aab9bb
  %[arXiv:1710.00845 [astro-ph.CO]].
  %%CITATION = doi:10.3847/1538-4357/aab9bb;%%
  
  %\cite{Aghanim:2018oex}
\bibitem{Aghanim:2018oex} 
  N.~Aghanim {\it et al.} [Planck Collaboration],
  %``Planck 2018 results. VIII. Gravitational lensing,''
  arXiv:1807.06210 [astro-ph.CO].
  %%CITATION = ARXIV:1807.06210;%%
  
  %\cite{Lewis:2002ah}
\bibitem{Lewis:2002ah} 
  A.~Lewis and S.~Bridle,
  %``Cosmological parameters from CMB and other data: A Monte Carlo approach,''
  Phys.\ Rev.\ D {\bf 66}, 103511 (2002).
  %doi:10.1103/PhysRevD.66.103511
  %[astro-ph/0205436].
  %%CITATION = doi:10.1103/PhysRevD.66.103511;%%
  
  \bibitem{Gelman-Rubin}
  A.~Gelman and D.~Rubin, 
  %{\it ``Inference from iterative simulation using multiple sequences,''} 
  Statistical Science \textbf{7}, 457 (1992).
  
  %\cite{Lewis:2013hha}
\bibitem{Lewis:2013hha} 
  A.~Lewis,
  %``Efficient sampling of fast and slow cosmological parameters,''
  Phys.\ Rev.\ D {\bf 87}, 103529 (2013).
  %doi:10.1103/PhysRevD.87.103529
  %[arXiv:1304.4473 [astro-ph.CO]].
  %%CITATION = doi:10.1103/PhysRevD.87.103529;%%
  
  %\cite{Asgari:2019fkq}
\bibitem{Asgari:2019fkq} 
  M.~Asgari {\it et al.},
  %``KiDS+VIKING-450 and DES-Y1 combined: Mitigating baryon feedback uncertainty with COSEBIs,''
  Astron.\ Astrophys.\  {\bf 634}, A127 (2020).
  %doi:10.1051/0004-6361/201936512
  %[arXiv:1910.05336 [astro-ph.CO]].
  %%CITATION = doi:10.1051/0004-6361/201936512;%%
  
\bibitem{Zarrouk:2018vwy}
P.~Zarrouk {\it et al.},
%``The clustering of the SDSS-IV extended Baryon Oscillation Spectroscopic Survey DR14 quasar sample: measurement of the growth rate of structure from the anisotropic correlation function between redshift 0.8 and 2.2,''
Mon. Not. Roy. Astron. Soc. \textbf{477}, 1639 (2018).
%doi:10.1093/mnras/sty506
%[arXiv:1801.03062 [astro-ph.CO]].


\bibitem{Agathe:2019vsu}
V.~de S. Agathe {\it et al.},
%``Baryon acoustic oscillations at z = 2.34 from the correlations of Ly$\alpha$ absorption in eBOSS DR14,''
Astron. Astrophys. \textbf{629}, A85 (2019).
%doi:10.1051/0004-6361/201935638
%[arXiv:1904.03400 [astro-ph.CO]].

\bibitem{Blomqvist:2019rah}
M.~Blomqvist {\it et al.}, 
%``Baryon acoustic oscillations from the cross-correlation of Ly$\alpha$ absorption and quasars in eBOSS DR14,''
Astron. Astrophys. \textbf{629}, A86 (2019).
%doi:10.1051/0004-6361/201935641
%[arXiv:1904.03430 [astro-ph.CO]].

%\cite{Heavens:2017hkr}
\bibitem{Heavens:2017hkr}
A.~Heavens, Y.~Fantaye, E.~Sellentin, H.~Eggers, Z.~Hosenie, S.~Kroon and A.~Mootoovaloo,
%``No evidence for extensions to the standard cosmological model,''
Phys. Rev. Lett. \textbf{119}, no.10, 101301 (2017)
doi:10.1103/PhysRevLett.119.101301
[arXiv:1704.03467 [astro-ph.CO]].

%\cite{Heavens:2017afc}
\bibitem{Heavens:2017afc}
A.~Heavens, Y.~Fantaye, A.~Mootoovaloo, H.~Eggers, Z.~Hosenie, S.~Kroon and E.~Sellentin,
%``Marginal Likelihoods from Monte Carlo Markov Chains,''
[arXiv:1704.03472 [stat.CO]].

%\cite{Kass:1995loi}
\bibitem{Kass:1995loi}
R.~E.~Kass and A.~E.~Raftery,
%``Bayes Factors,''
J. Am. Statist. Assoc. \textbf{90}, no.430, 773-795 (1995)
doi:10.1080/01621459.1995.10476572

\end{thebibliography}
\end{document}